\documentclass[a4paper,preprint,11pt,number]{elsarticle}

\usepackage{amsmath,amssymb}
\usepackage{booktabs}
\usepackage{float}
\usepackage{longtable}
\usepackage{CJKutf8}
\usepackage[hidelinks]{hyperref}

\begin{document}
\begin{frontmatter}

\title{Uncovering Latent Patterns in Program-Associated Library Attendance through Background Adjustment}

\author[imi]{Keunsu Kim\corref{corresponding}}
\ead{kim.keunsu.752@m.kyushu-u.ac.jp}
\author[imi]{Hisatoshi Kodani\corref{corresponding}}
\ead{kodani@imi.kyushu-u.ac.jp}
\author[innovation]{Motoki Masada}
\author[engineering]{Nam Tuan Nguyen}

\address[imi]{Institute of Mathematics for Industry, Kyushu University, Fukuoka, Japan}
\address[innovation]{Joint Graduate School of Mathematics for Innovation, Kyushu University, Fukuoka, Japan}
\address[engineering]{Department of Civil Engineering, Graduate School of Engineering, Kyushu University, Fukuoka, Japan}

\cortext[corresponding]{Corresponding authors. The authors are listed in alphabetical order by family name.}

\begin{abstract}
Routinely collected library entrance records offer an opportunity to understand attendance patterns associated with library programs, but ordinary daily variation and detailed administrative user status categories complicate their interpretation. Building on a project-based learning collaboration with Kyushu University Central Library, we analyzed entrance records from October 2018 to July 2026. Descriptive analysis revealed relatively modest differences in attendance across the main weather categories, higher attendance on days with mean temperatures of 30°C or above, and pronounced differences between weekdays and weekends. To account for such background variation, we estimated expected attendance for each user status category and applied nonnegative matrix factorization (NMF) to identify common patterns in positive background-adjusted
responses from 92 events across 39 user status categories. Three interpretable response modes were identified: two concentrated on temporary external visitors and undergraduates, respectively, and one representing a broader combination of graduate student and research-related categories. The three-mode solution was selected by jointly considering held-out reconstruction error and basis stability across year-excluded fits. The identified modes also remained stable after excluding events from the selected COVID-19 period. The modes appeared in different combinations across individual events, with their dominance varying by program category. These findings illustrate how existing library records can reveal interpretable patterns of program-associated attendance without predefined grouping of user status categories.
\end{abstract}

\end{frontmatter}

\section{Introduction}
\label{sec:introduction}

This study originated from a project-based learning (PBL)
collaboration with Kyushu University Central Library.
The project sought ways to use routinely collected
entrance records and program information to better
understand patterns of library use.
In addition to total daily attendance, the library
records entrance counts separately for detailed administrative user status categories. These data provide an opportunity to examine
how different types of users attend the library during
periods associated with library programs.

Two challenges must be considered when analyzing
program-associated attendance patterns using these data.
First, daily library attendance can vary with ordinary
background conditions such as day of the week, the
academic calendar, opening hours, and weather conditions.
Attendance observed during a particular program therefore
cannot be interpreted directly as a response to that program.
Previous studies have reported periodic patterns in
academic library circulation associated with weekdays
and the academic calendar~\cite{mcgrath1996periodicity}.
Daily public library circulation has also been examined
in relation to conditions such as day of the week,
holidays, and precipitation~\cite{sato2021effect}.
These studies support the need to account for ordinary
background variation before examining
program-associated changes in attendance.

Second, the administrative user status categories used
by the library do not necessarily represent distinct groups
in terms of their responses to library programs.
After data cleaning, 39 user status categories remained available for analysis. One possible approach would be to combine these categories into broader groups, such as students, faculty and staff, and external users. Although this would simplify the analysis, it would require deciding in advance which user status categories to group together, making it difficult to explore common response patterns directly from the data. Conversely, analyzing all 39 categories independently would make it difficult to summarize common response patterns across categories. We therefore sought a data-driven approach to identify a small number of shared response patterns without requiring predefined grouping of the original user status categories.

To address these challenges, this study connects the
exploratory analysis conducted during the PBL project
with a subsequent analysis of program-associated responses.
First, the students participating in the PBL project
examined the distributions of daily attendance across
weather categories, daily mean temperature ranges,
and days of the week.
This descriptive analysis characterized how observed
attendance varied across ordinary background conditions
and provided a starting point for interpreting
program-associated changes in attendance.

We then estimated the attendance expected for each user status category under the observed background conditions and constructed program-specific responses from the differences between observed and expected attendance on each program's verified activity dates. To summarize and interpret these responses, we sought to represent patterns across user status categories using a small number of interpretable latent response modes and to examine how these modes were combined within individual programs. Focusing on patterns of higher-than-expected attendance, we therefore applied nonnegative matrix factorization (NMF) to the positive parts of the program-specific responses. NMF provides a low-rank representation of nonnegative data and has been widely used to extract interpretable latent patterns \cite{lee1999learning,lee2000algorithms}. Unlike principal component analysis (PCA), which allows both positive and negative contributions, NMF represents each program's positive response as an additive combination of nonnegative modes with nonnegative coefficients. This prevents cancellation between modes and allows their contributions within individual programs to be interpreted separately.

The choice of NMF rank directly affects both
the number and the interpretation of the extracted
response modes. Factorizations at different ranks
do not necessarily have a nested component structure
\cite{zhang2015nested}, so improved reconstruction
alone is insufficient to determine an appropriate
rank. We therefore evaluated candidate ranks
using two complementary criteria. First, we evaluated how accurately the response modes estimated from the remaining years could reconstruct the responses of events from the excluded year. Second, we examined whether the estimated response modes remained similar to those obtained from the full dataset when events from one year were excluded. Together, these assessments allowed us to consider both held-out reconstruction error and the stability of the response modes when selecting the NMF rank.

Together, these analyses characterize ordinary variation in daily library attendance and examine whether increases associated with programs, after accounting for background conditions, can be summarized by a small number of stable response modes across user status categories.
We also examine how these modes are distributed across five categories of library programs.

The descriptive analysis showed substantial overlap
in attendance distributions across weather categories,
while days with mean temperatures of $30^\circ\mathrm{C}$
or above had higher mean and median attendance than
the other temperature groups.
Differences across days of the week were particularly
pronounced, especially between weekdays and weekends.
The subsequent NMF analysis identified three positive response modes: one dominated by temporary external visitors, another by undergraduates, and a third characterized by a broader mixture of graduate student and research-related categories.
The distributions of these modes varied across program categories, demonstrating how routinely collected entrance records can be used to characterize program-associated attendance patterns.


\section{Data and Descriptive Background Variation}
\label{sec:data}

\subsection{Library attendance and program records}
\label{sec:library_data}

The study used daily entrance records from Kyushu University
Central Library from October 1, 2018, when the new library
fully opened on the Ito Campus, through July 31, 2026.
Each entrance was recorded together with an administrative
user status label.
The original data contained 45 administrative user status labels.
Five labels with zero recorded entrances and one invalid administrative code were excluded, leaving 39 user status categories for analysis.
The complete list is provided in Appendix \hyperref[app:user_status]{A}.

Program records were obtained from the
Kyushu University Library event listings
and included event names, recorded start
and end dates, information URLs, and venues.
The 187 original records were reviewed
for Central Library relevance, program type,
and identifiable activity dates.
Cancellation or postponement notices and
records outside the target program categories
were excluded.
Exhibitions were excluded from the
target event analysis because their recorded
periods often extended over many days.
Instead, exhibition activity was included
as a covariate in the baseline model described in Section~\ref{sec:baseline}.
Programs without verifiable activity dates
on open-library days were also excluded.
After applying these criteria and treating
multiple records of the same event as one,
92 events were retained for analysis.
The records did not contain event-specific
participant counts.
For each selected event, verified activity
dates coinciding with Central Library open
days were used rather than the entire
recorded start--end interval.
Table~\ref{tab:event_categories} summarizes
the selected events across five program categories.

\par\medskip
\noindent
\begin{minipage}{\linewidth}
\centering

\refstepcounter{table}
\label{tab:event_categories}

\small
\textbf{Table~\thetable.}
Categories of library events included in the analysis.

\medskip

\setlength{\tabcolsep}{8pt}

\begin{tabular}{@{}lr@{}}
\toprule
Program category & Events \\
\midrule
Seminar / Workshop & 33 \\
Participation Program & 43 \\
Lecture / Symposium & 11 \\
Open Campus & 3 \\
Library Tour & 2 \\
\midrule
\textbf{Total} & \textbf{92} \\
\bottomrule
\end{tabular}

\end{minipage}
\par\medskip

The library calendar additionally provided daily opening
information, including opening schedule type and opening hours.
Daily opening status was determined from the official calendar.
Across the study period, 2,602 days were identified as valid
open-library days.

These records were used both to characterize ordinary variation
in daily attendance and, in the subsequent analysis, to estimate
the attendance expected for each user status category before
constructing program-specific responses.

\subsection{Weather data}
\label{sec:weather_data}

Daily weather data for Fukuoka were obtained from the historical
weather records of the Japan Meteorological Agency
(JMA)~\cite{jma_weather_data}.
We used daily mean temperature and the daytime weather summary
reported for 06:00--18:00.

The JMA weather summary consists of a basic weather term followed,
when applicable, by terms describing subsequent or temporary changes.
Because the first basic term represents the weather condition
prevailing for the largest portion of the observation period,
each day was classified according to this first term.

The observed basic weather terms were grouped into four broad
categories: Sunny, Cloudy, Rainy, and Snowy.
The original Japanese terms and their corresponding categories
are listed in Appendix \hyperref[app:weather_terms]{B}.
Days indicating mixed snow and rain were included in the
Snowy category.

\subsection{Descriptive variation in daily attendance}
\label{sec:descriptive_background}

Before constructing program-specific responses, we examined
how daily attendance varied across weather conditions,
daily mean temperature, and days of the week.
Because Open Campus attracts unusually large numbers of
visitors and is concentrated in the summer, its associated
dates could disproportionately influence these descriptive
comparisons, particularly those involving temperature.
We therefore excluded 11 open days associated with Open Campus,
leaving 2,591 days for all three comparisons.

Daily mean temperature was grouped into four ranges,
\[
\begin{aligned}
T<10^\circ\mathrm{C},\qquad
&10\leq T<20^\circ\mathrm{C},\\
20\leq T<30^\circ\mathrm{C},\qquad
&T\geq30^\circ\mathrm{C}.
\end{aligned}
\]

\begin{figure}[H]
    \centering
    \includegraphics[width=\linewidth]
    {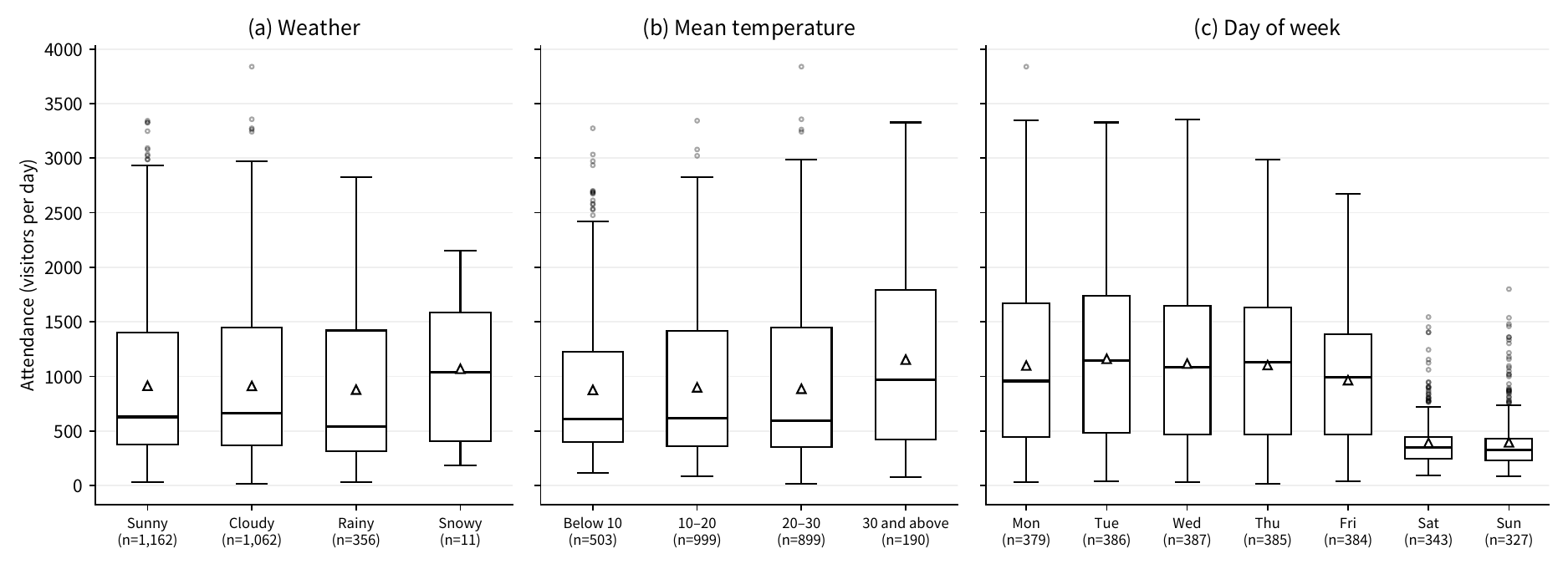}
    \caption{
   Distribution of daily attendance across ordinary background
conditions.
Panels show attendance by (a) daytime weather category,
(b) daily mean temperature range, and (c) day of the week.
All panels use the same 2,591 days on which the library was open,
after excluding 11 days associated with Open Campus.
Boxes indicate the interquartile range, horizontal lines
indicate medians, and triangles indicate means.
Whiskers extend to the most extreme observations within
1.5 times the interquartile range from the first and third
quartiles, and points indicate observations beyond the whiskers.
    }
    \label{fig:background_variation}
\end{figure}

Figure~\ref{fig:background_variation} summarizes the attendance
distributions across weather categories, temperature ranges,
and days of the week.
Table~\ref{tab:background_variation} reports the corresponding
numbers of days, means, medians, and standard deviations.


\begin{table}[!htbp]
\centering
\begin{minipage}{\linewidth}
\centering

\refstepcounter{table}
\label{tab:background_variation}

\small
\textbf{Table~\thetable.}
Daily attendance across weather conditions,
mean temperature ranges, and days of the week
($N=2{,}591$).

\medskip

\small
\setlength{\tabcolsep}{4pt}

\begin{tabular}{@{}lrrrr@{}}
\toprule
Group & Days & Mean & Median & SD \\
\midrule
\multicolumn{5}{l}{\textit{Weather condition}} \\
Sunny  & 1,162 & 913.8 & 628.0 & 679.3 \\
Cloudy & 1,062 & 912.8 & 659.5 & 650.5 \\
Rainy   & 356 & 878.6 & 538.5 & 695.9 \\
Snowy   & 11 & 1,069.6 & 1,037.0 & 708.4 \\
\addlinespace
\multicolumn{5}{l}{\textit{Daily mean temperature ($^\circ\mathrm{C}$)}} \\
Below 10 & 503 & 876.3 & 608.0 & 644.6 \\
10--20 & 999 & 899.6 & 616.0 & 633.8 \\
20--30 & 899 & 886.5 & 591.0 & 669.3 \\
30 and above & 190 & 1,154.0 & 970.5 & 854.4 \\
\addlinespace
\multicolumn{5}{l}{\textit{Day of week}} \\
Mon & 379 & 1,099.4 & 958.0 & 727.6 \\
Tue & 386 & 1,161.5 & 1,147.5 & 728.1 \\
Wed & 387 & 1,116.9 & 1,083.0 & 695.6 \\
Thu & 385 & 1,103.1 & 1,129.0 & 662.5 \\
Fri & 384 & 965.0 & 990.0 & 545.0 \\
Sat & 343 & 391.0 & 348.0 & 220.8 \\
Sun & 327 & 394.8 & 329.0 & 265.8 \\
\bottomrule
\end{tabular}

\end{minipage}
\end{table}


Among the 2,591 days, 1,162 were classified as sunny,
1,062 as cloudy, 356 as rainy, and 11 as snowy.
Mean attendance was 913.8, 912.8, 878.6, and 1,069.6
visitors per day, respectively.
Attendance distributions overlapped substantially across
the major weather categories, with relatively small
differences in their central tendencies.
Although the Snowy category had a higher mean and median,
it contained only 11 days and should therefore be
interpreted cautiously.

A different pattern was observed across temperature ranges.
Mean attendance in the three ranges below $30^\circ\mathrm{C}$
was similar, at 876.3, 899.6, and 886.5 visitors per day,
respectively.
In contrast, days with mean temperatures of
$30^\circ\mathrm{C}$ or above had a mean attendance of
1,154.0 visitors.
This was approximately 28.3\% higher than the mean of
899.6 visitors in the $10$--$20^\circ\mathrm{C}$ group.
The corresponding medians were 970.5 and 616.0 visitors,
respectively, a difference of approximately 57.5\%.
The higher attendance in the hottest range was therefore evident in both measures of central tendency.
The pattern was not monotonic across temperature ranges,
as the three groups below $30^\circ\mathrm{C}$ showed
broadly similar attendance levels.

The largest differences were observed across days of the week.
Mean attendance from Monday through Thursday ranged from
1,099.4 to 1,161.5 visitors per day, whereas Friday
had a somewhat lower mean of 965.0 visitors.
Attendance was substantially lower on weekends, with
means of 391.0 visitors on Saturday and 394.8 on Sunday.
The contrast between weekdays and weekends was particularly
pronounced compared with the differences observed across
weather categories and temperature ranges.

Taken together, these descriptive results show substantial
overlap in attendance distributions across weather categories,
higher mean and median attendance in the highest temperature
range, and particularly large differences between weekdays
and weekends.
Although these descriptive comparisons do not establish
causal relationships, they highlight the need to account
for ordinary background variation when examining
program-associated attendance.
The following section develops a background-adjusted
analysis of program-specific attendance responses.


\section{Analysis}
\label{sec:analysis}

Section~\ref{sec:descriptive_background} showed that daily library attendance
varied across ordinary calendar and environmental conditions. We therefore
first estimated the attendance expected for each user status category under
these background conditions and then quantified program-associated deviations
from those expectations. Because the present study focuses on
higher-than-expected attendance, the positive parts of these
background-adjusted responses were subsequently summarized using nonnegative
matrix factorization.

\subsection{Background-adjusted baseline attendance}
\label{sec:baseline}

Let $C_{ds}$ denote the observed entrance count for user status $s$ on open day $d$. To reduce the influence of unusually large counts and accommodate days with zero attendance, we applied the log transformation
\begin{equation}
    Y_{ds}=\log(1+C_{ds}).
    \label{eq:log_count}
\end{equation}
Separate baseline models were fitted for each of the 39 user status categories using 1,839 open-library days outside primary-event activity dates and the recorded periods of programs with unverified activity dates. For each status $s$,
\begin{equation}
Y_{ds}
=
\beta_{0s}
+
\mathbf{x}_d^\top\boldsymbol{\beta}_s
+
\varepsilon_{ds},
\label{eq:baseline_model}
\end{equation}
where $\beta_{0s}$ is the intercept,
$\boldsymbol{\beta}_s$ is the coefficient vector,
and $\varepsilon_{ds}$ is the error term.

The predictor vector $\mathbf{x}_d$ included
day of the week, ISO week, fiscal year,
opening schedule type, opening hours,
exhibition activity, and daily mean temperature.
Day of the week, ISO week, fiscal year, and
opening schedule type were represented by
dummy variables, while opening hours and
daily mean temperature were treated as
continuous variables and exhibition activity
as a binary indicator.

Daily mean temperature was retained as an
environmental covariate, whereas the broad
weather categories, whose attendance distributions
overlapped substantially in the descriptive
analysis, were not included in the baseline model.

The regression coefficients were estimated by ordinary least squares, yielding fitted values $\widehat{Y}_{ds}$ and residuals $\widehat{\varepsilon}_{ds}=Y_{ds}-\widehat{Y}_{ds}$. To avoid underestimating expected attendance when converting back from the log scale, we used Duan's nonparametric smearing estimator \cite{duan1983smearing}. For each user status $s$, a correction factor was calculated from the residuals as
\begin{equation}
    S_s
    =
    \frac{1}{|D_0|}
    \sum_{d\in D_0}
    \exp(\widehat{\varepsilon}_{ds}),
    \label{eq:smearing_factor}
\end{equation}
where $D_0$ denotes these 1,839 baseline training days.

The fitted values were then converted back to the original count scale using this correction factor:
\begin{equation}
    \widehat{C}_{ds}
    =
    \max
    \left\{
        S_s\exp(\widehat{Y}_{ds})-1,
        0
    \right\}.
    \label{eq:expected_count}
\end{equation}
The maximum ensures that the estimated attendance is nonnegative. Thus, $\widehat{C}_{ds}$ represents the expected attendance for user status $s$ on day $d$ after accounting for the background conditions.

\subsection{Construction of event-specific responses}
\label{sec:event_response}

For event $e$, let $D_e$ denote the verified activity dates
on which the Central Library was open.
For each user status $s$, we defined the background-adjusted
response as the average difference between observed and
expected daily attendance over these dates:
\begin{equation}
    A_{es}
    =
    \frac{1}{|D_e|}
    \sum_{d\in D_e}
    \left(
        C_{ds}-\widehat{C}_{ds}
    \right).
    \label{eq:event_response}
\end{equation}

A positive value of $A_{es}$ indicates that attendance for
status $s$ was higher than expected on average during event
$e$, whereas a negative value indicates lower-than-expected
attendance. This daily average allows events of different
durations to be compared on the same scale.

Because event activity dates may overlap, the same open day
can contribute to more than one event response. Such
responses describe attendance associated with event activity
dates; they do not identify event participants or separate
causal effects of concurrent programs.

Collecting the responses across the 92 events and
39 retained user status categories gives the
signed event-response matrix
\begin{equation}
    A=(A_{es})
    \in
    \mathbb{R}^{92\times39}.
    \label{eq:signed_A}
\end{equation}

\subsection{NMF representation of positive status-specific responses}
\label{sec:nmf_representation}

The signed event-response matrix $A$ contains both positive and negative
deviations from expected attendance. Because this study focuses on
higher-than-expected attendance, only the positive responses were used
for NMF:
\begin{equation}
    A^+_{es}
    =
    \max(A_{es},0).
    \label{eq:positive_part}
\end{equation}
The NMF input matrix was therefore
\begin{equation}
    X
    =
    A^+
    \in
    \mathbb{R}_+^{92\times39}.
    \label{eq:nmf_input}
\end{equation}
For a given rank $K$, NMF approximates
\begin{equation}
    X
    \approx
    WV,
    \qquad
    W\in\mathbb{R}_+^{92\times K},
    \qquad
    V\in\mathbb{R}_+^{K\times39}.
    \label{eq:nmf}
\end{equation}
For event $e$,
\begin{equation}
    X_e
    \approx
    \sum_{k=1}^{K}
    W_{ek}V_k,
    \label{eq:nmf_event}
\end{equation}
where $V_k$ represents the $k$th response mode across the 39 user status
categories and $W_{ek}$ is its coefficient for event $e$.
Unlike PCA, which allows positive and negative contributions that may
cancel each other, NMF represents each event response as an additive
combination of nonnegative modes.

The factors were estimated by minimizing generalized Kullback--Leibler
divergence, a standard objective for NMF \cite{lee2000algorithms}:
\begin{equation}
    (\widehat{W},\widehat{V})
    =
    \arg\min_{W,V\geq0}
    D_{\mathrm{KL}}(X\|WV).
    \label{eq:nmf_objective}
\end{equation}
Each NMF model was fitted using one NNDSVDa initialization and 30 random
initializations, and the solution with the lowest generalized KL
divergence was retained.

After fitting, each basis vector was normalized so that
\begin{equation}
    \sum_{j=1}^{39}V_{kj}=1.
    \label{eq:basis_normalization}
\end{equation}
The corresponding column of $W$ was rescaled so that $WV$ remained unchanged.
Since all entries are nonnegative, $V_{ks}$ can be interpreted as the
relative contribution of user status category $s$ within mode $k$, while $W_{ek}$
represents the magnitude of that mode's contribution to event $e$.

\subsection{Rank selection and stability}
\label{sec:rank_selection}

We considered candidate ranks $K=2,\ldots,7$.
In PCA, increasing the rank adds new components while
preserving the preceding ones, so rank selection determines
how much of the data variation is retained.
In contrast, NMF does not generally produce nested components:
increasing the rank can change existing response modes
rather than simply adding new ones \cite{zhang2015nested}.

Previous studies have proposed NMF rank-selection criteria
based on the stability of sample assignments or basis components
across different random initializations
\cite{brunet2004metagenes,kim2007sparse, kotliar2019identifying}.
These approaches rely on the idea that well-defined patterns
should be consistently recovered across repeated runs.
In this study, we sought to determine whether the identified
response modes depended strongly on events from particular years.
We therefore evaluated each candidate rank using reconstruction
error for events held out by year and the stability of the
basis vectors when events from individual years were excluded.

The 92 events were grouped by their start year.
Let $\mathcal Y$ denote the set of event start years,
and let $X^{(y)}$ and $X^{(-y)}$ denote the
submatrices of $X$ containing events from year $y$
and all other years, respectively.
For each year $y$, rank-$K$ NMF was fitted to
$X^{(-y)}$:
\begin{equation}
X^{(-y)}
\approx
W^{(-y,K)}V^{(-y,K)}.
\end{equation}
The resulting basis vectors were used for two assessments.

\begin{enumerate}
    \item \textbf{Held-out reconstruction.}
    The basis $V^{(-y,K)}$ was held fixed,
    and the nonnegative coefficients were estimated
    by minimizing generalized KL divergence:
    \begin{equation}
    \widehat W^{(y,K)}
    =
    \arg\min_{W\geq 0}
    D_{\mathrm{KL}}
    \left(
    X^{(y)}\middle\|WV^{(-y,K)}
    \right).
    \end{equation}
    The excluded events were then reconstructed as
    \begin{equation}
    \widehat X^{(y,K)}
    =
    \widehat W^{(y,K)}V^{(-y,K)}.
    \end{equation}
    The total reconstruction error was
    \begin{equation}
    E_K=
    \sum_{y\in\mathcal Y}
    D_{\mathrm{KL}}
    \left(
    X^{(y)}\middle\|\widehat X^{(y,K)}
    \right).
    \end{equation}
    Each of the 92 events contributed once
    to the total error.

    \item \textbf{Basis stability.}
    Using the same yearly fits, we compared
    $V^{(-y,K)}$ with the full-data basis $V^{(K)}$.
    Basis vectors were matched one-to-one by maximizing
    cosine similarity.
    Let $s_{y,k}$ denote the cosine similarity between
    the matched basis vectors for mode $k$ when year $y$
    was excluded.
    We summarized stability by averaging these similarities
    over all years and modes:
    \begin{equation}
    S_K=
    \frac{1}{K|\mathcal Y|}
    \sum_{y\in\mathcal Y}\sum_{k=1}^{K}s_{y,k}.
    \end{equation}
\end{enumerate}

\begin{figure}[H]
    \centering
    \includegraphics[width=\linewidth]
    {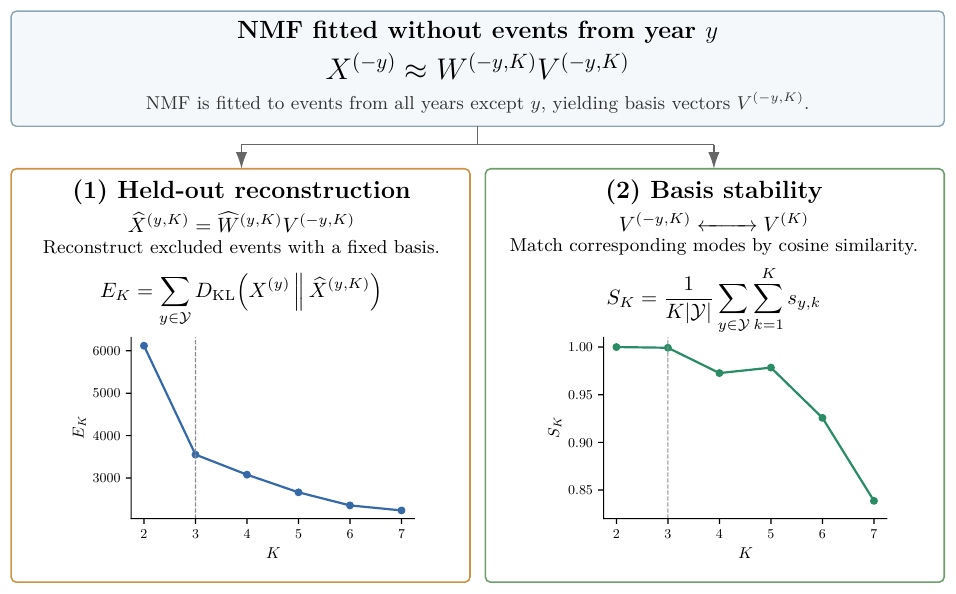}
    \caption{
    Rank selection for positive-only KL-NMF
    ($X=A^+\in\mathbb{R}_+^{92\times39}$).
    For each event start year, NMF was fitted using
    events from all other years.
    Panel (1) shows the total generalized KL divergence
    from reconstructing the excluded events with
    the fixed basis.
    Panel (2) shows the mean cosine similarity between
    matched year-excluded and full-data basis vectors,
    averaged over all years and modes.
    Dashed vertical lines indicate the selected rank $K=3$.
    }
    \label{fig:rank_selection}
\end{figure}

Figure~\ref{fig:rank_selection} summarizes the two assessments.
The held-out KL divergence decreased by 41.9\%
from $K=2$ to $K=3$, compared with a further
13.3\% reduction from $K=3$ to $K=4$.
Meanwhile, the mean basis cosine similarity remained
high at $K=3$ ($S_3=0.9992$), but decreased
to $S_4=0.9727$.
Thus, $K=3$ provided a substantial improvement in held-out reconstruction accuracy while maintaining high basis stability.
We therefore selected $K=3$ as a parsimonious
balance between reconstruction accuracy and
basis stability.


\subsection{Positive response modes and program-level patterns}

\begin{figure}[H]
    \centering
    \includegraphics[width=\linewidth]
    {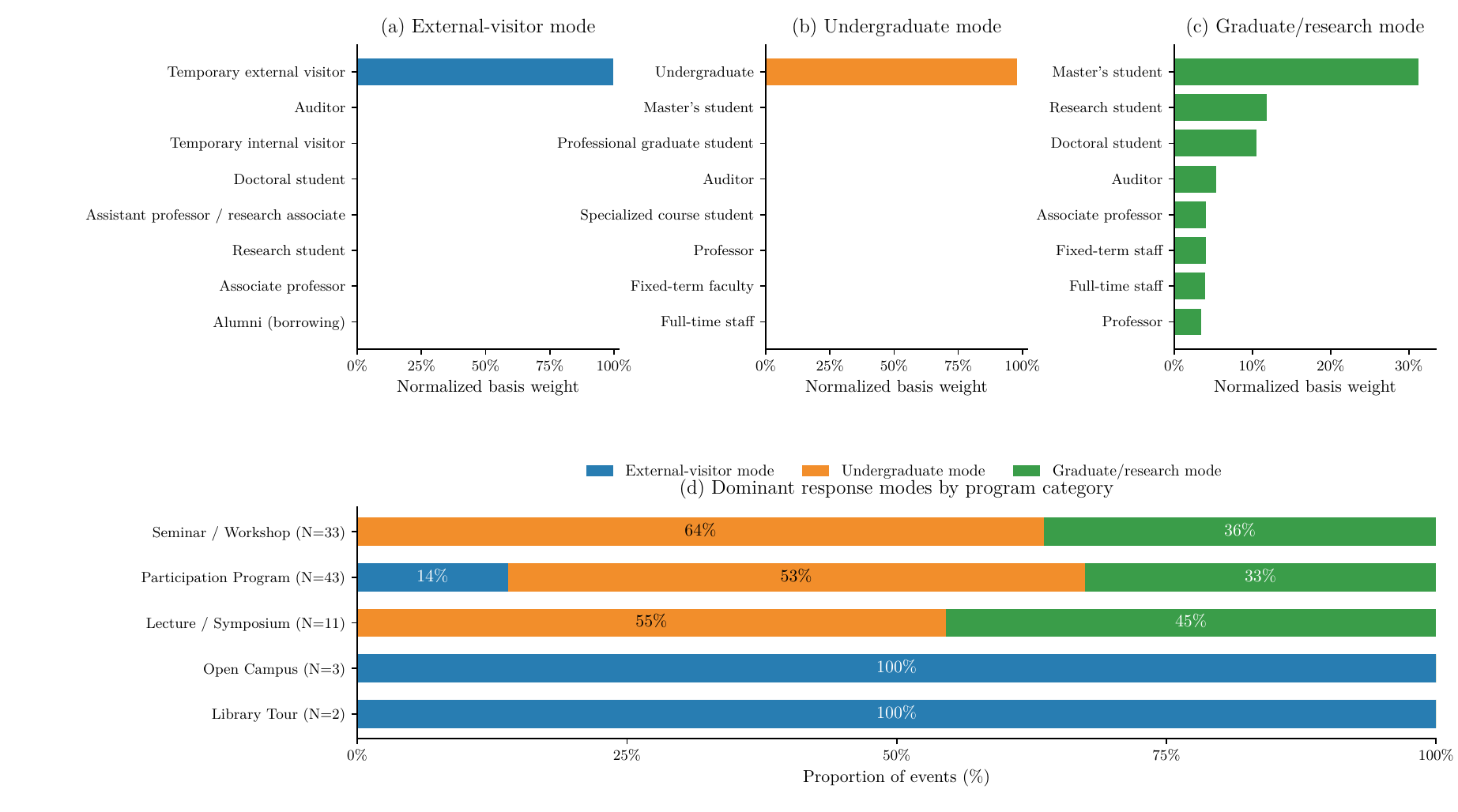}
    \caption{
    Positive response modes and their distributions
    across program categories for the selected
    $K=3$ solution.
    Panels (a)--(c) show the eight largest normalized
    basis weights for the external-visitor,
    undergraduate, and graduate/research modes,
    respectively.
    Panel (d) shows the proportions of events assigned
to each dominant mode within the five program
categories, using all 92 events.
    The dominant mode of each event was determined by the largest normalized mode coefficient $p_{ek}$ defined in Eq.~\eqref{eq:mode_share}, using the full-data NMF solution.
    }
    \label{fig:response_modes}
\end{figure}

The selected $K=3$ solution identified three positive
response modes. Figures~\ref{fig:response_modes}(a)--(c)
show the eight largest normalized basis weights
for each mode. The modes were obtained without
predefined user status groupings and were named
according to their principal characteristics.

\noindent\textbf{External-visitor mode.}
Temporary external visitors accounted for 99.6\%
of the normalized basis weight.
This mode was therefore strongly concentrated
in temporary external visitors.

\noindent\textbf{Undergraduate mode.}
Undergraduates accounted for 97.8\% of the
normalized basis weight.
All remaining categories had smaller contributions.

\noindent\textbf{Graduate/research mode.}
The third mode was more broadly distributed
across user status categories. Master's students
had the largest weight (31.2\%), followed by
research students (11.8\%), doctoral students
(10.5\%), auditors (5.4\%), and associate
professors (4.0\%).

To compare the relative contributions of
response modes within each event, we normalized
the mode coefficients $W_{ek}$ introduced
in Section~\ref{sec:nmf_representation}:
\begin{equation}
p_{ek}
=
\frac{W_{ek}}{\sum_{\ell=1}^{K}W_{e\ell}}.
\label{eq:mode_share}
\end{equation}
Because each basis vector was normalized to
sum to one, $p_{ek}$ represents the relative
contribution of mode $k$ to the reconstructed
positive response of event $e$.
The mode with the largest value of \(p_{ek}\) was referred to as the dominant mode.

Across all 92 events, the undergraduate mode
was dominant in 50 events (54.3\%), followed
by the graduate/research mode in 31 events
(33.7\%) and the external-visitor mode in
11 events (12.0\%).

Figure~\ref{fig:response_modes}(d) summarizes the distribution of
dominant modes across the five program categories
listed in Table~\ref{tab:event_categories}.
Among the 33 seminars and workshops,
the undergraduate mode was dominant in 21 events
and the graduate/research mode in 12.
Similarly, among the 43 participation programs,
23 were dominated by the undergraduate mode,
14 by the graduate/research mode, and six
by the external-visitor mode.
Among the 11 lectures and symposia,
six were dominated by the undergraduate mode
and five by the graduate/research mode.
In contrast, all three Open Campus events
and both library tours were dominated by
the external-visitor mode.
These results show that the distributions of dominant modes differ across program categories.

However, the dominant mode represents only
the largest contribution to each event,
and other modes may also contribute.
To illustrate these contributions,
Figures~\ref{fig:event_mode_C1}--\ref{fig:event_mode_C3}
in \ref{app:event_level_modes}
show the mode coefficient shares of all 92 events,
grouped by program category. These event-level views complement the
dominant-mode distributions
in Figure~\ref{fig:response_modes}(d) by showing
the variation in mode composition within
and across program categories.

\paragraph{Sensitivity to the COVID-19 period}
As an additional sensitivity check, we applied
the basis-matching approach used in Section~\ref{sec:rank_selection}
after excluding 16 events whose activity dates
overlapped the period from April 1, 2020, to
May 7, 2023. This period spans the start of the
2020 academic year through the withdrawal of
Kyushu University's COVID-19 action guidelines
\cite{kyushu2023guidelines}.
The NMF basis vectors at rank $K=3$ obtained from
the remaining 76 events closely matched those
from the full dataset, with a mean cosine
similarity of 0.9998 across the three matched
modes. The dominant-mode assignments of all
76 retained events were also unchanged.


\section{Conclusion}
\label{sec:conclusion}

This study examined how routinely collected library
entrance records can be used to characterize attendance
patterns associated with library programs.
Building on a PBL collaboration
with Kyushu University Central Library, we combined
descriptive analysis of daily attendance with an analysis
of program-associated responses after accounting for
ordinary background variation.
The descriptive results showed substantial overlap
in attendance distributions across weather categories,
higher attendance on days with mean temperatures of
$30^\circ\mathrm{C}$ or above, and particularly pronounced
differences between weekdays and weekends.
These findings highlighted the importance of considering
ordinary background conditions when examining attendance
during library programs.

Using NMF,
we identified three interpretable patterns of
higher-than-expected attendance across 92 events
and 39 administrative user status categories.
Two patterns were strongly concentrated in single
user status categories, corresponding to temporary
external visitors and undergraduates, whereas the
third captured a broader combination of graduate student
and research-related categories.
The three patterns remained consistent when events
from individual years were excluded, and the three-mode
representation reconstructed held-out event responses
more accurately than the two-mode representation.
This suggests that the identified patterns are not
limited to a particular period and may be useful
for understanding library attendance responses
across different years.
Similar results were also obtained after excluding
events overlapping the selected COVID-19 period.
Furthermore, individual events could exhibit
contributions from multiple patterns, and the
distribution of dominant patterns differed
across program categories.

These findings demonstrate that complex attendance
records can be summarized into a small number of
meaningful patterns without predefined aggregation
of administrative user status categories.
The analysis identified both responses concentrated
in particular user groups and responses distributed
across several groups, while showing how these
patterns were expressed across different library programs.
Although the identified patterns describe associations
between attendance and program activity dates
rather than causal effects of individual programs,
the approach provides a practical way to better
understand library attendance using existing
institutional data.
More broadly, this study illustrates how an initial
student-led investigation through PBL can be extended
into a systematic mathematical analysis of real-world
data.

\paragraph{Acknowledgements}
The collaborative course described in this paper was made possible by the generous support of Kyushu University Central Library, including the provision of data and assistance with classroom activities. We would like to express our sincere gratitude to Kyushu University Central Library for their valuable contributions. We would also like to thank Professor Shuichi Kawano and Professor Hayato Waki for their guidance and support throughout this project.

\begin{CJK*}{UTF8}{ipxm}
\bibliographystyle{elsarticle-num}
\bibliography{references}

@article{duan1983smearing,
  author  = {Duan, Naihua},
  title   = {Smearing Estimate: A Nonparametric Retransformation Method},
  journal = {Journal of the American Statistical Association},
  volume  = {78},
  number  = {383},
  pages   = {605--610},
  year    = {1983},
  doi     = {10.1080/01621459.1983.10478017}
}

@article{mcgrath1996periodicity,
  author  = {McGrath, William E.},
  title   = {Periodicity in academic library circulation: A spectral analysis},
  journal = {Journal of the American Society for Information Science},
  volume  = {47},
  number  = {2},
  pages   = {136--145},
  year    = {1996}
}

@article{sato2021effect,
  author  = {Sato, Sho and Harada, Takashi and Tatebe, Kanako
             and Izumi, Yui and Suemichi, Ayano and Inaba, Momoka
             and Igarashi, Tomoya and Koizumi, Masanori},
  title   = {Effect of Weather and Days of the Week on Number of Daily Circulation in Public Libraries},
  journal = {{LIBRES}: Library \& Information Science Research Electronic Journal},
  volume  = {31},
  number  = {1},
  pages   = {56--65},
  year    = {2021},
  doi     = {10.32655/LIBRES.2021.1.5}
}

@article{lee1999learning,
  author  = {Lee, Daniel D. and Seung, H. Sebastian},
  title   = {Learning the parts of objects by non-negative matrix factorization},
  journal = {Nature},
  volume  = {401},
  number  = {6755},
  pages   = {788--791},
  year    = {1999},
  doi     = {10.1038/44565}
}

@inproceedings{lee2000algorithms,
  author    = {Lee, Daniel D. and Seung, H. Sebastian},
  title     = {Algorithms for Non-negative Matrix Factorization},
  booktitle = {Advances in Neural Information Processing Systems},
  volume    = {13},
  pages     = {556--562},
  year      = {2000}
}

@article{brunet2004metagenes,
  author  = {Brunet, Jean-Philippe and Tamayo, Pablo
             and Golub, Todd R. and Mesirov, Jill P.},
  title   = {Metagenes and molecular pattern discovery using matrix factorization},
  journal = {Proceedings of the National Academy of Sciences},
  volume  = {101},
  number  = {12},
  pages   = {4164--4169},
  year    = {2004},
  doi     = {10.1073/pnas.0308531101}
}

@article{zhang2015nested,
  author  = {Zhang, Lingsong and Lu, Shu and Marron, J. S.},
  title   = {Nested nonnegative cone analysis},
  journal = {Computational Statistics \& Data Analysis},
  volume  = {88},
  pages   = {100--110},
  year    = {2015},
  doi     = {10.1016/j.csda.2015.01.008}
}

@misc{jma_weather_data,
  author = {{Japan Meteorological Agency}},
  title  = {Historical Weather Data Search: Fukuoka},
  note   = {Daily meteorological observations.
            Accessed October 2026.
            \url{https://www.data.jma.go.jp/stats/etrn/index.php}}
}

@article{kotliar2019identifying,
  author  = {Kotliar, Dylan and Veres, Adrian and Nagy, M. Aurel
             and Tabrizi, Shervin and Hodis, Eran
             and Melton, Douglas A. and Sabeti, Pardis C.},
  title   = {Identifying gene expression programs of cell-type identity and cellular activity with single-cell {RNA-Seq}},
  journal = {eLife},
  volume  = {8},
  pages   = {e43803},
  year    = {2019},
  doi     = {10.7554/eLife.43803}
}

@misc{kyushu2023guidelines,
  author = {{Kyushu University, Student Support Division}},
  title  = {{課外活動の制限解除について（通知）}},
  year   = {2023},
  month  = may,
  note   = {May 16, 2023.
            \url{https://www.kyushu-u.ac.jp/f/52764/kagaikatudou20230516.pdf}}
}

@article{kim2007sparse,
  author  = {Kim, Hyunsoo and Park, Haesun},
  title   = {Sparse non-negative matrix factorizations via alternating non-negativity-constrained least squares for microarray data analysis},
  journal = {Bioinformatics},
  volume  = {23},
  number  = {12},
  pages   = {1495--1502},
  year    = {2007},
  doi     = {10.1093/bioinformatics/btm134}
}
\end{CJK*}

\clearpage


\appendix
\setcounter{table}{0}
\setcounter{figure}{0}
\section{User Status Categories}
\label{app:user_status}

\begin{CJK}{UTF8}{ipxm}

The raw attendance data contained 45 administrative user status labels.
Five categories---\textit{Other} (その他), \textit{External medical
institution} (学外医療機関), \textit{Special use} (特別利用),
\textit{Reciprocal use} (相互利用), and \textit{Relocation work}
(移転作業)---had zero recorded entrances across the entire study period
and were excluded. The label \textit{Status code requiring correction}
(要身分コード修正) was also excluded because it did not represent a valid
user status category. This left 39 user status categories for analysis.

Table~\ref{tab:user_status_categories} lists the original Japanese labels
and the corresponding English labels for these 39 retained categories.
The categories were retained individually and were not combined into broader
groups before NMF.

\end{CJK}

\begin{CJK}{UTF8}{ipxm}

\begin{longtable}{p{0.34\textwidth}p{0.58\textwidth}}
\caption{Original Japanese user status categories and the corresponding English labels used in the analysis.}
\label{tab:user_status_categories}\\

\hline
Original Japanese label & English label \\
\hline
\endfirsthead

\hline
Original Japanese label & English label \\
\hline
\endhead

※区分外
& Unclassified \\

その他学生
& Other student \\

一般市民(入館専用)
& General public (entry only) \\

一般市民(貸出可)
& General public (borrowing) \\

他大学学生等(入館専用)
& Other-university student (entry only) \\

他大学学生等(貸出可)
& Other-university student (borrowing) \\

他大学教職員(入館専用)
& Other-university faculty/staff (entry only) \\

他大学教職員(貸出可)
& Other-university faculty/staff (borrowing) \\

元教員
& Former faculty \\

准教授
& Associate professor \\

助教・准助教
& Assistant professor / research associate \\

医員
& Medical staff (physician) \\

卒業生(入館専用)
& Alumni (entry only) \\

卒業生(病院地区)
& Alumni (hospital campus) \\

卒業生(貸出可)
& Alumni (borrowing) \\

名誉教授
& Professor emeritus \\

大学院生(修士)
& Master's student \\

大学院生(博士)
& Doctoral student \\

大学院生(専門職)
& Professional graduate student \\

学外各種学校(入館専用)
& Other-school student (entry only) \\

学外各種学校(貸出可)
& Other-school student (borrowing) \\

学部生
& Undergraduate \\

専修生
& Specialized course student \\

放送大学学生
& Open University student \\

教務職員
& Academic affairs staff \\

教員(特別・有期)
& Fixed-term faculty \\

教授
& Professor \\

特別研究員
& Research fellow \\

研修医
& Resident physician \\

研究員
& Researcher \\

研究生
& Research student \\

科目等履修生
& Non-degree course student \\

聴講生
& Auditor \\

職員(常勤)
& Full-time staff \\

職員(特別・有期)
& Fixed-term staff \\

臨時入館専用(学内者)
& Temporary internal visitor \\

臨時入館専用(学外者)
& Temporary external visitor \\

講師
& Lecturer \\

非常勤講師
& Part-time lecturer \\

\hline

\end{longtable}

\end{CJK}


\section{Japanese Weather Terms}
\label{app:weather_terms}

The weather categories used in the descriptive analysis were based on the
first basic weather term in the JMA daytime weather summary for 06:00--18:00.
The Japanese terms observed in the study period were mapped to the four broad
categories as follows.

\begin{CJK}{UTF8}{ipxm}

\begin{center}
\begin{tabular}{lll}
\hline
Japanese term & English meaning & Category \\
\hline
快晴   & Clear skies                   & Sunny  \\
晴     & Sunny weather                      & Sunny  \\
曇     & Cloudy weather                      & Cloudy \\
薄曇   & Thin cloud cover                   & Cloudy \\
雨     & Rain                         & Rainy   \\
大雨   & Heavy rain                   & Rainy   \\
雪     & Snow                         & Snowy  \\
みぞれ & Sleet / mixed snow and rain & Snowy   \\
\hline
\end{tabular}
\end{center}

\end{CJK}


\clearpage

\section{Event-Level Response-Mode Composition}
\label{app:event_level_modes}

\begin{figure}[H]
    \centering
    \includegraphics[width=\linewidth]{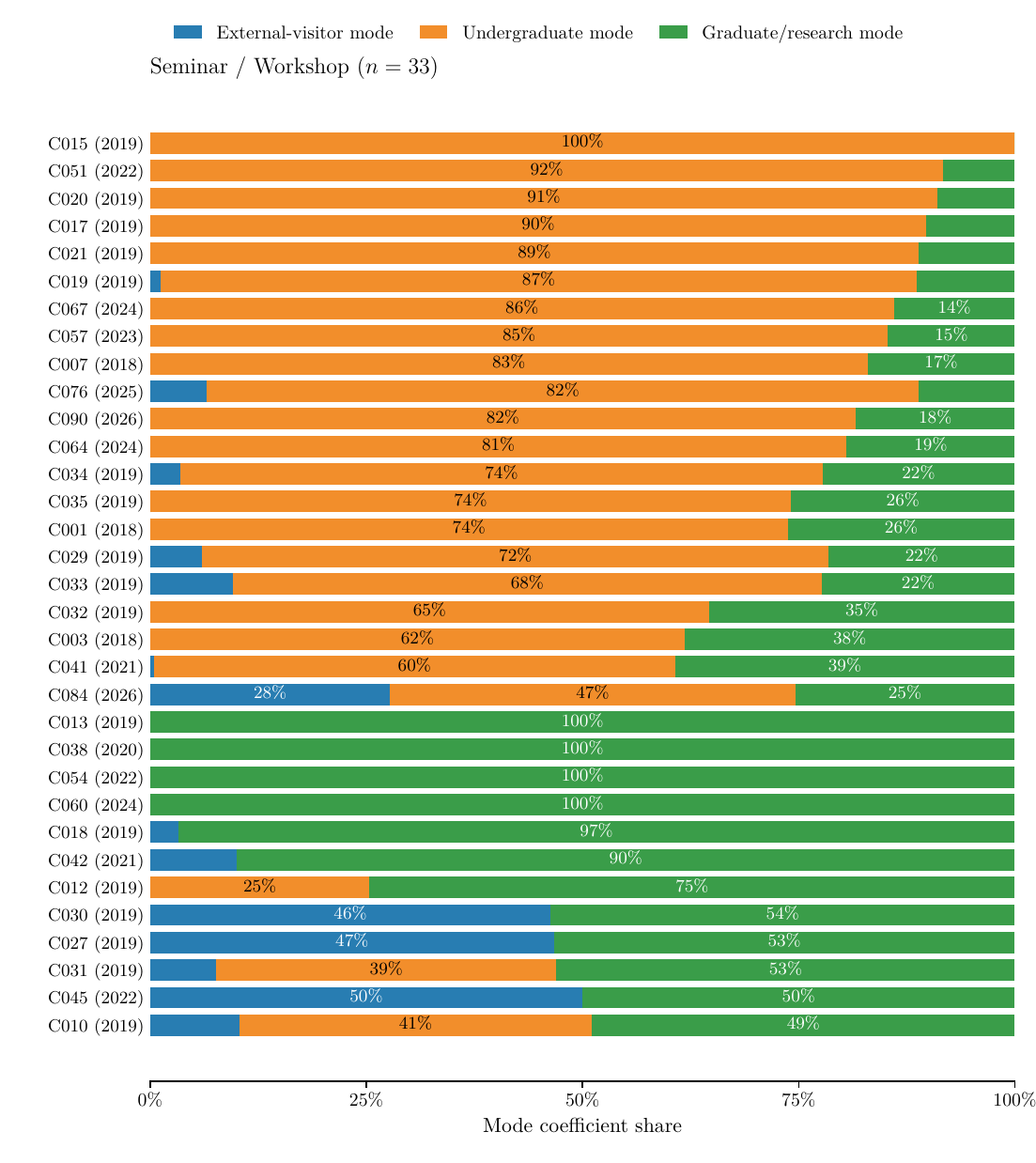}

    \caption{
    Event-level response-mode composition for 33 seminar
and workshop events.
    The stacked bars show coefficient shares from the
    full-data $K=3$ positive-only NMF solution.
    The three shares sum to 100\% within each event and
    represent contributions to the reconstructed positive
    response, not proportions of participants.
    Labels are shown for shares of at least 13\%.
    Events are grouped by their dominant mode and
sorted within each group by decreasing
dominant-mode share, with event ID used to break ties.
    }
    \label{fig:event_mode_C1}
\end{figure}

\begin{figure}[H]
    \centering
    \includegraphics[width=\linewidth]{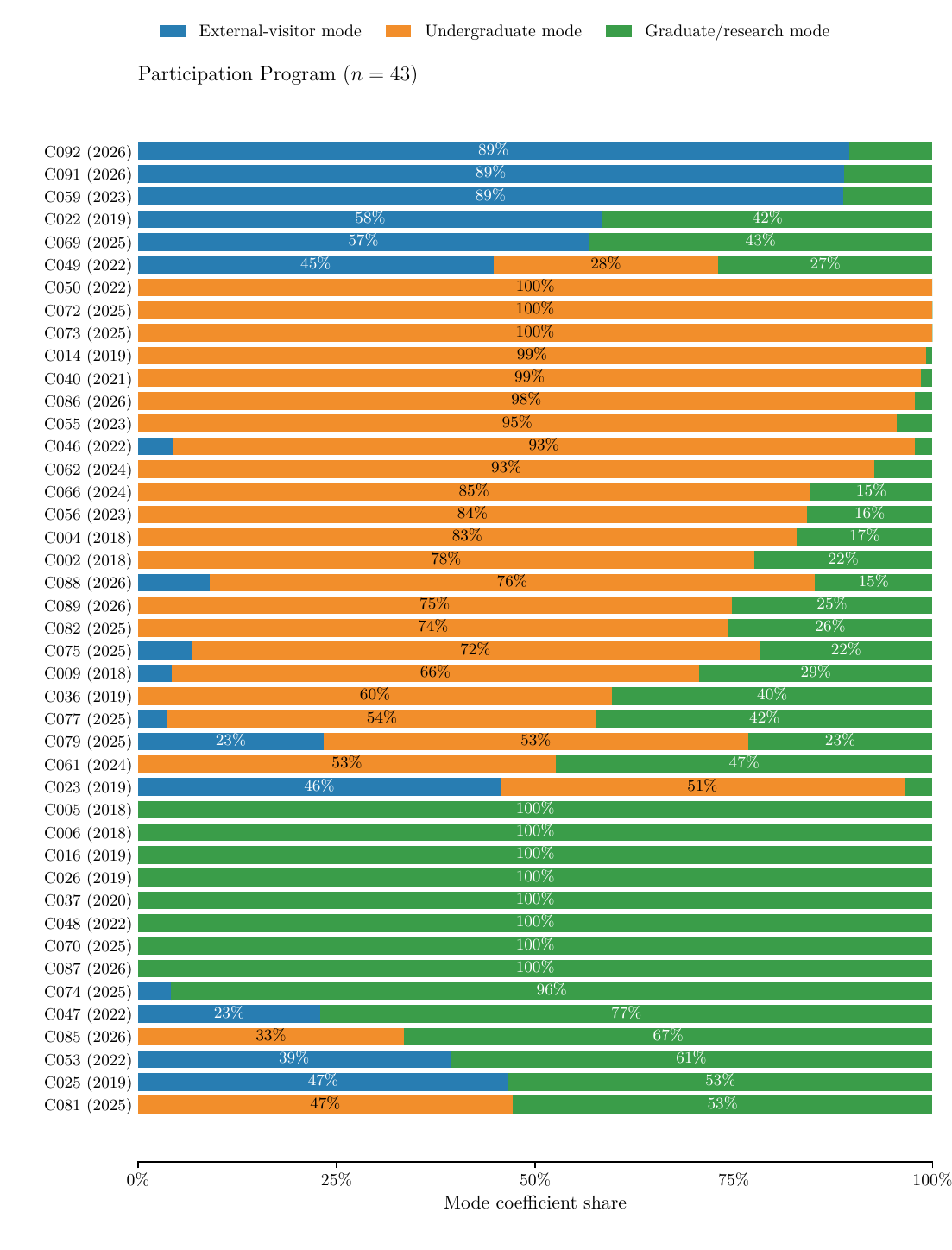}

    \caption{
    Event-level response-mode composition for 43
    participation-program events.
    Coefficient shares are obtained from the same
    full-data $K=3$ positive-only NMF solution.
    The shares represent relative contributions to the
    reconstructed positive response, not participant
    proportions.
    Labels are shown for shares of at least 13\%.
   Events are grouped by dominant mode and sorted
within each group by decreasing dominant-mode share,
with event ID used to break ties.
    }
    \label{fig:event_mode_C2}
\end{figure}

\begin{figure}[H]
    \centering
    \includegraphics[width=\linewidth]{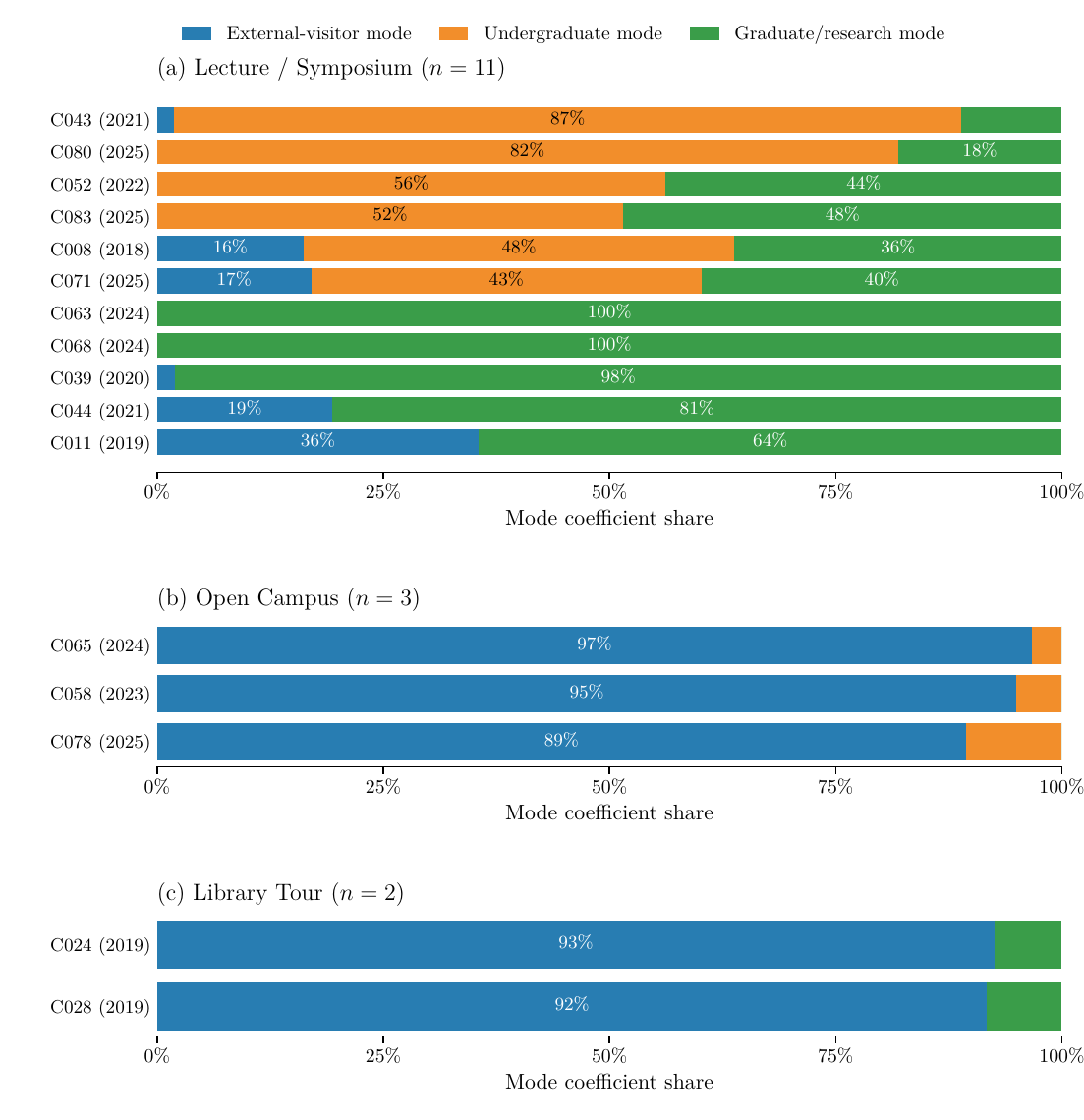}

    \caption{
    Event-level response-mode composition for
11 lectures and symposia, 3 Open Campus events,
and 2 library tours.
    Coefficient shares are obtained from the same
    full-data $K=3$ positive-only NMF solution.
    The shares represent relative contributions to the
    reconstructed positive response, not participant
    proportions.
    Labels are shown for shares of at least 13\%.
Within each program category, events are grouped
by dominant mode and sorted within each group
by decreasing dominant-mode share, with event ID
used to break ties.
    }
    \label{fig:event_mode_C3}
\end{figure}

Figures~\ref{fig:event_mode_C1}--\ref{fig:event_mode_C3}
show the mode coefficient shares $p_{ek}$ defined in
Eq.~\eqref{eq:mode_share} for all 92 library events,
computed from the full-data $K=3$ positive-only NMF
solution and grouped by program category.


\clearpage


\clearpage


\end{document}